# Classification of Attacks in Wireless Sensor Networks

Mohamed-Lamine Messai
Doctoral school in computer science, university of Bejaia, Algeria
Networks & Distributed Systems Laboratory, UFAS, Setif, Algeria
messai.amine@gmail.com

**Abstract— In wireless sensor networks (WSNs), security has a vital importance. Recently, there was a huge interest to propose security solutions in WSNs because of their applications in both civilian and military domains. Adversaries can launch different types of attacks, and cryptography is used to countering these attacks. In this paper, we present challenges of security, and classification of the different possible attacks in WSNs. The problems of security in each layer of the network's OSI model are discussed.**

## I. INTRODUCTION

Rapid development in technologies realizations of electronic components, and in particular, of microprocessors, make possible to develop equipment with low cost and low power, of size and weights increasingly reduced. WSNs are a particular type of ad hoc networks, comprised mainly of large number (hundred or thousand) deployed sensor nodes with limited resources and one or more base stations (BSs) or sink (Figure 1), typically serves as the access point for the user or as a gateway to another network. Nodes can collect and transmit (with wireless links) environmental data (temperature, pressure, humidity, noise levels, etc) in autonomous manner. The node in WSN plays tow roles: collect data and route data back to the base station.

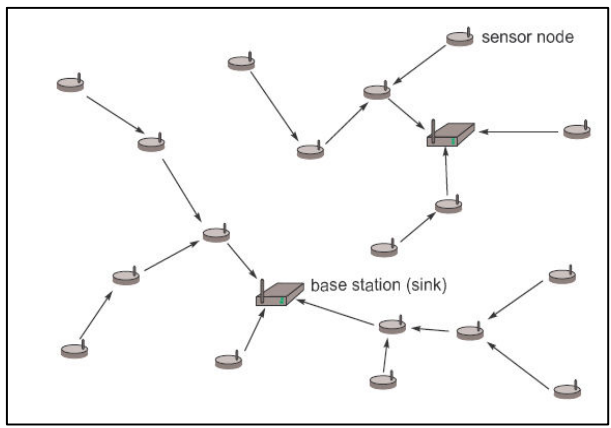

Figure 1. A WSN.

Typically, sensor node consists of five components, as shown in figure 2: power unit (battery), memory, transmitter/receiver, embedded processor, and sensing unit. Additional components can be implanted in a sensor node: location finding system: allow the node to find its position, a power generator: used for recharging battery node and prolong its lifetime, and a mobilizer: make nodes move [8]. Sensing units are usually composed of two subunits: sensor and analog-to-digital converter (ADC). When an event was produced (analog data) node sense analog signal observed convert it to digital signals by the ADC unit, and then treat it with the processing unit. A transceiver unit connects the node to the network. One of the most important components of a sensor node is the power unit (battery); it is the fuel of the node. For detailed stat of the art see [8, 15].

For example, SmartDust node have only 8 bit processor, an 8 KB instruction flash memory, and a bandwidth of 10 Kbps (Kilo bit per second) [12].

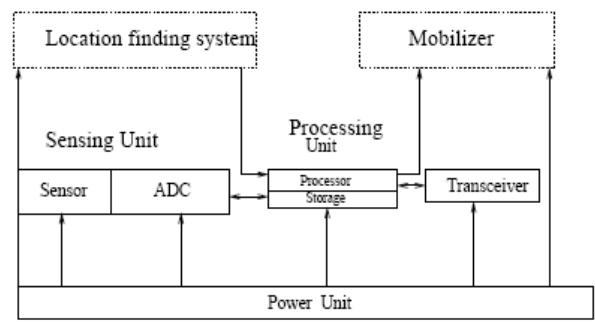

Figure 2. Sensor node [8].

As much of technology, the development of WSNs was caused by military needs. Indeed, the armies discreetly wish to be in measurement of espionner their enemies. SOSUS (the SOund SUrveillance System) used during the cold war to detect Soviet submarines [13, 14], other applications are: environmental monitoring, health monitoring of patients, habitat monitoring, disaster recovery, smart environment.

Sensor nodes are deployed in hostile and inaccessible area (e.g. military use to enemy surveillance), and it is impossible (in general) to know the position of the node after deployment, so nodes can be physically captured or destroyed by attackers. Provide security is a very important problem in WSNs.

Traditional security techniques used in traditional networks can not be applied directly, because of extremely constrained resources like energy, bandwidth and capabilities of processing and storing data of nodes in WSNs. The tiny hardware of sensor node is not capable of performing complex security protocols, so security should be reconsidered and new ideas in security researches are needed.

In this paper, we aim to give an overview of security problem in WSNs, present different attacks, and classify these attacks in the OSI model. The reminder of this paper is organized as follows: in section 2, OSI model of these networks is briefly presented. Sections 3-6 discuss constraints and limitation of sensor nodes and security goals. Sections 7-9 give definitions of attacks, attackers, and impact of attacks toward each layer of OSI model. Finally, we conclude in section 10.

## II. THE OSI MODEL

The role of this model (Opening System Interconnect) consists in standardizing the communication between the participants so that various manufacturers can develop compatible products (software or hardware).

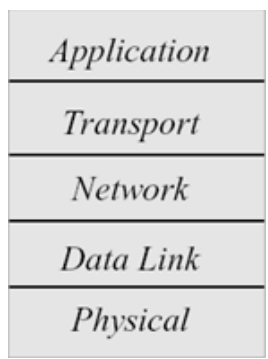

Figure 3.  OSI model in WSNs.

Each layer of the model communicates with an adjacent layer (that of the top or that of the lower part). Each layer uses the services of the sub-bases and provides some to that of higher level.

## III. VULNERABILITY ANALYSIS

The physical vulnerability is the fact that sensors are scattered in insecure place like public places, the natural environments (mountainous region) as well as the buildings, smart houses and museums (smart environment), so attacker have the physical access to the node, and with appropriate tools, he can read the secret information (like keys, programs, etc) stored in the node memory.

Other vulnerability is related to wireless technology. Unlike the traditional wired networks, the attackers could easily capture the data packet because the data transmissions are all in the air. Whoever having the adequate receiver can potentially listen to or disturb the exchanged messages.

Sensor nodes are themselves routers. Packets pass through different nodes in multi-hops routes to arrive at their destination. Due to the possible of violation of such nodes, this feature presents a serious vulnerability.

Sensor nodes are prone to failure, witch make topology dynamic. Dynamic network topology can be caused also by the mobility of nodes and addition of new nodes.

## IV. CONSTRAINTS INFLUENCING ON SECURITY IN WSNs

Constraints that make traditional security impractical in WSNs are:

Low energy power: The energy of the nodes is limited (limited battery lifetime), and generally irreplaceable and no-recharging battery. Protocols of WSNs must concentrate mainly on the conservation of energy.

Limited memory and computation capacity: In the majority of WSNs, nodes are not able to memorize keys of significant size, or to carry out complex protocols cryptographic [1].

Therefore, new security measures are needed to address constraints of WSNs.

## V. ENERGY FOR SECURITY

Lifetime node is generally limited by the lifetime of a tiny battery, so energy is the fundamental resource constraint. The additional power consumed by nodes of sensor due to security is dependent on:

- Calculation necessary for the functions of security, such as ciphering, deciphering, verification of the signature.
- Energy necessary for the transmission and management of security material (keys, etc).
- Energy necessary for the storage of the keys.

The challenge is to minimize the consumption of the energy with maximizing the performances of security.

Energy is an important factor to consider when designing security measures for WSNs. Conserving node energy to extend his lifetime, and prolong network functionalities.

## VI. SECURITY GOALS

Sensor networks with limited processing power, storage, bandwidth, and energy require special security approaches. The hardware and energy constraints of the sensors add difficulty to the security requirements of ad hoc networks concerning availability, integrity, confidentiality, freshness, authentication, access control, and non-repudiation [2].

*Availability*: the availability gives insurance over the reactivity and time of response of the system to transmit information of one source to the good destination. It also means that the services of network are available to the authorized parts if necessary and ensures the services of network in spite of denies of service attack (DoS).

*Integrity*: it is a service which guarantees that data are not be modified during the transmission. Integrity protects the network against the injection or the modification of messages.

*Confidentiality*: is the guarantee that the information of a node is not available or revealed only with its recipient.

*Freshness*: WSNs provide some measurements in time; we must ensure that each message is fresh. The freshness of data implies that the data are recent, and it ensures that no adversary replay the old messages.

*Authentication*: an adversary is not simply limited to modify the message. He can inject additional messages. Thus the receiver must make sure that the data used come from the correct source. In addition, by constructing WSNs, the authentication is necessary for many tasks.

*Access control*: gives to the legitimate participants a means to detect the messages coming from external sources of the network.

*Non-repudiation*: ensures that the origin of a message cannot deny having sent the message [2].

VII. ATTACKS IN WSNs

A variety of attacks against WSNs is documented in the literature. To face these attacks, various against measurements were proposed. We present in the continuation the principal types of attacks, and in section 9 we assign these attacks to the layers concerned of the OSI model.

A classification of the attacks consists in distinguishing the passive attacks from the active attacks.

The passive attack (eavesdropping) is limited to listening and analyzes exchanged traffic. This type of attacks is easier to realize (it is enough to have the adequate receiver), and it is difficult to detect. Since, the attacker does not make any modification on exchanged information. The intention of the attacker can be the knowledge of confidential information or the knowledge of the significant nodes in the network (cluster head node), by analyzing routing information, to prepare an active attack.

In the active attacks, an attacker tries to remove or modify the messages transmitted on the network. He can also inject his own traffic or replay of old messages to disturb the operation of the network or to cause a denial of service. Among the most known active attacks, we can quote:

*Tampering*: it is the result of physical access to the node by an attacker; the purpose will be to recover cryptographic material like the keys used for ciphering [3].

*Black hole*: a node falsifies routing information to force the passage of the data by itself, later on; its only mission is then, nothing to transfer, creating a sink or black hole in the network [1].

*Selective forwarding*: as mentioned above, a node play the role of router, in a selective forwarding attack, malicious nodes may refuse to forward certain messages and simply drop them.

*Sybil attack*: Newsome et al. [5] definite this attack by: "malevolent device, taking multiple identities in an illegitimate way", attacker can use the identities of the others nodes in order to take part in distributed algorithms such as the election.

*HELLO flood attack*: many routing protocols use "HELLO" packet to discover neighboring nodes and thus to establish a topology of the network. The simplest attack for an attacker consists in sending a flood of such messages to flood the network and to prevent other messages from being exchanged.

*Jamming*: a well-known attack on wireless communication, it consists in disturbing the radio channel by sending useless information on the frequency band used. This jamming can be temporary, intermittent or permanent [6].

*Blackmail attack*: a malicious node makes announce that another legitimate node is malicious to eliminate this last from the network. If the malicious node manages to tackle a significant number of nodes, it will be able to disturb the operation of the network.

*Exhaustion*: is to consume all the resources energy of the victim node, by obliging it to do calculations or to receive or transmit unnecessarily data [7].

*Wormhole attack*: attackers here are strategically placed at different ends of a network. They can receive messages and replays them in different parts by means of a tunnel [9].

*Identity replication attack*: attacker can clone nodes, and place it in different part of the network in order to collect majority of information traffic. Unlike the Sybil attack, the identity replication attack [10] is based upon giving the same identity to different physical nods. This attack can be mounted because in a WSN there is no way to know that a wireless sensor node is compromised.

VIII. ATTACKER MODEL

The goal of an attacker (adversary) is to illegally obtain keys stored in nodes by vulnerabilities exploitation.

Strong attacker: The adversary is considered as present before and after deployment of nodes. It can supervise all the communications, anywhere, and at any moment.

A realistic attacker model: The attacker is able to supervise a fixed percentage of communication channels after deployment [4].

*"The hostile surveillance is not ubiquitous during the deployment phase of the network and only fraction of the established link keys can be obtained by the attacker"* [4].

## IX. PROBLEMS OF SECURITY IN EACH LAYER

In this section, we provide a layer based classification of defined attacks on the OSI model described above.

### A. Physical layer

Deal with the specification of the frequencies bands. This layer must ensure of the techniques of emission, reception and modulation of data in a robust way.

The attacks associated in the physical layer are very few but, at the same time, can be most difficult to prevent: jamming on the same frequency that the network uses, and the physical attack of a node.

One standard defence against jamming employs spread-spectrum communication [7]. Node capture is the more upsetting problem in the security of WSNs. The use of resistant hardware against capture attack (tamperproof node) can solve this problem, but increase the node cost.

In figure 5, we use MATLAB software (MATLAB is an abbreviation of MATrix LABoratory. MATLAB is an interactive matrix-based system for scientific and engineering calculations) to simulate a random deployment of 100 sensor nodes in 100m x 100m area (figure 4), and then we randomly compromise sensor nodes to evaluate the effect of compromised links.

Whenever the number of neighbors of the compromised sensor node is high, the great is the effect of compromised this sensor node.

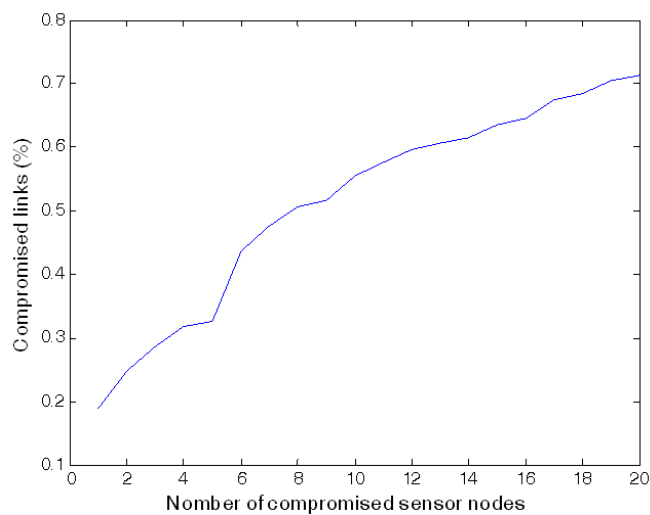

Figure 5. Percentage of compromised links versus number of compromised sensor nodes.

### B. MAC layer

This layer manages the access to the radio channel (MAC layer), and control errors. The adversary can only induce collision in a one byte of a transmission to disturb the entire data packet. So obligate the victim node to retransmit the data packet and cause a death of this node by consuming its energy (exhaustion attack).

The prevention of these attacks can be limited to impose the use of small packets, use techniques of correction to ask for the retransmission of packet.

### C. Network layer

WSNs use a communication multi-hops for routing the packets towards the destination, the attacks in this layer are: black hole, selective forwarding, Sybil attack, HELLO flood attack, wormhole, and Identity replication attack.

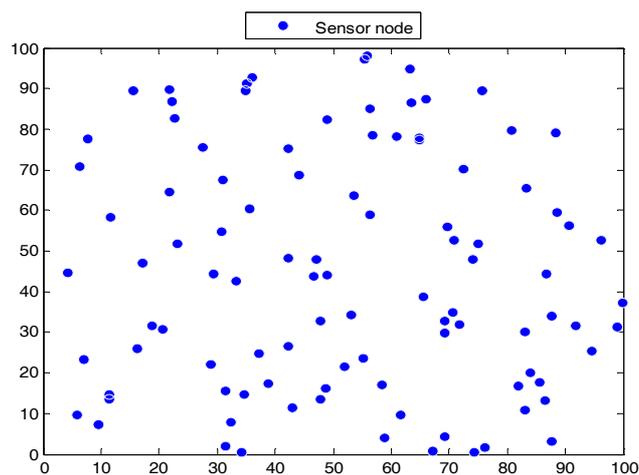

Figure 4. A WSN with 100 sensor nodes.

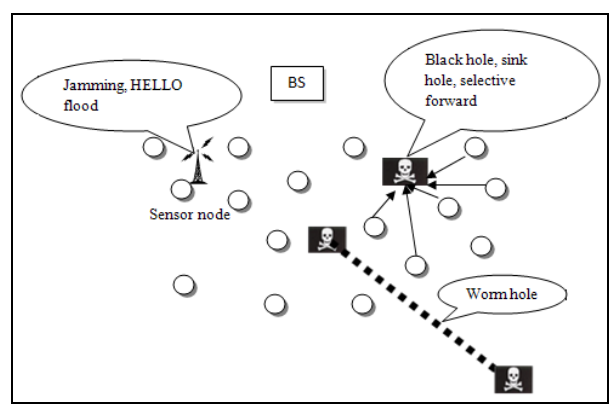

Figure 6. Routing layer attacks in WSNs.

The prevention of this kind of attacks invites to authenticate all messages. In a hierarchical network of sensors, the parents' nodes can check the identity of the source of a packet in transit.

Monitoring is a strategy to make safe the routing and to detect the abnormal behaviors of the nodes [11]. In this approach, the nodes act as "a watchdogs" to supervise the next transmission of the packet. If the misbehaviour would be detected, the nodes will update the information of routing to avoid the compromised node.

Probing is another proactive defense against the malevolent nodes in WSNs [11]. This method periodically sends packets of probing through the network to detect areas of breakdown. Since the geographical protocols of routing have the knowledge of the topology of the network, probing is particularly well adapted to their use. The probing packets must seem to be the normal traffic, in order to detect the compromised nodes.

Redundancy another approach suggested in [11] consists in sending the package several times on different paths; at least a path delivers the packet to the destination. It is clear that this method does not preserve energy but it increases the difficulty for an attacker of stopping the traffic.

X. DISCUSSION

Securing WSNs is a subject of active work. There are three issues to underwrite security in WSNs; (i) key management: to use encryption, the parties involved have to hold the right cryptographic keys. Key management schemes are essentials for every system to provide confidentiality, integrity, authentication, and the other security goals [16]. It is the technique of establishment and maintenance of keys between legitimate nodes, and allows the updating, revocation and destruction of keys. Due to the resource limitation, providing efficient key management in WSNs is a challenge. (ii) Securing routing is the next issue to address. There are two kinds of threats to the routing protocols: external attackers, compromised interns' nodes, which are very difficult to detect because the compromised node can generate valid packets. Existents routing protocols for WSNs offer little or no security features [1]. (iii) Prevention of denial-of-service is the third issue, DoS can be defined as any event which decreases or eliminates the capacity of the network to carry out the functions envisaged. Breakdowns of hardware, programming errors, exhaustion of resource, environmental conditions, or any complicated interaction between these factors can cause DoS. DoS attacks prevent or reduce the use of computer or resources, interrupt or delay services, making network become unavailable, isolate legitimate users from a network.

XI. CONCLUSION

Now, popularity of WSNs increases, and takes attention of many researchers. This paper treats security challenges in WSNs, which differ from the ad hoc networks with more severe restrictions in terms of energy, computation capabilities and communications. Consequently, the solutions of security must thus be adapted.